# Microsecond-Pulsed Nanocalorimetry: A Scalable Approach for Ultrasensitive Heat Capacity Measurements

*Hugo Gómez-Torres[1†], Manel Molina-Ruiz[2†^], Simone Privitera[2], Enric Menéndez[2], Llibertat Abad[3], Jordi Sort[1,4], Olivier Bourgeois[5], Javier Rodriguez-Viejo[1,2], Aitor Lopeandia[1,2*]*

[1] Catalan Institute of Nanoscience and Nanotechnology (ICN2), CSIC and BIST, E-08193 Bellaterra, Spain.
[2] Departament de Física, Universitat Autònoma de Barcelona, E-08193 Bellaterra, Spain.
[3] Institut de Microelectrònica de Barcelona (IMB-CNM-CSIC). Campus de la UAB, E-08193, Cerdanyola del Vallès (Barcelona), Spain.
[4] Institució Catalana de Recerca i Estudis Avançats (ICREA), Pg. Lluís Companys 23, E-08010 Barcelona, Spain.
[5] Institut Neel, CNRS et Université Joseph Fourier, 25 avenue des Martyrs, F-38042 Grenoble Cedex 9, France

[†] *Both authors contributed equally to this work*
[^] Current address: Department of Physics, University of California, Berkeley, California 94720, USA
*e-mail: aitor.lopeandia@uab.cat

We introduce a nanocalorimetric technique based on microsecond-pulsed heating (μs-PHnC) that enables high-sensitivity, quasi-isothermal heat capacity measurements on nanoscale samples. Such resolution is critical for exploring thermodynamic signatures in low-dimensional materials, where conventional techniques fall short. By confining thermal excitation to microsecond timescales, this approach minimizes lateral heat diffusion, reduces heat capacity addenda to below $10^{-9}$ $J\cdot K^{-1}$, and achieves noise densities as low as 75 $pJ\cdot K^{-1}\cdot\sqrt{Hz}\cdot mm^{-2}$, unlocking precise thermodynamic characterization of subnanogram samples in areas as small as $30 \times 30$ $\mu m^2$. The method delivers exceptional temperature homogeneity, as demonstrated by resolving sharp phase transitions, such as the antiferromagnetic transition in ultrathin CoO films, with unprecedented clarity. Its quasi-static operation is inherently compatible with external stimuli, including magnetic and electric fields, thereby expanding its utility for in-operando thermodynamic studies. This advancement establishes a robust and scalable platform for probing thermal phenomena in nanostructured and low-dimensional materials, significantly broadening the scope of nanocalorimetry.

## 1. Introduction

The scaling down and production of materials with nanometric dimensions has revealed profound size-dependent effects in their thermodynamic behavior, including shifts in phase transition temperatures, altered enthalpic landscapes, and modified heat capacities[1,2]. These phenomena are particularly relevant in ultrathin films, nanostructures, and low-dimensional systems, where surface-to-volume ratios are high and energy exchanges are minimal. Accessing such thermodynamic quantities—heat capacity, enthalpy, and entropy changes—requires calorimetric techniques capable of resolving minute thermal signals. However, conventional calorimetry, which relies on bulk sample masses to ensure sufficient signal-to-noise ratio (SNR), is inherently limited in this context. The typical sample requirements and thermal inertia of traditional setups render them unsuitable for nanomaterials, which are often available only in nanogram quantities and exhibit low-energy transitions that are easily masked by background noise[3,4].

Chip calorimetry has emerged as a powerful technique to overcome the limitations of conventional calorimetry, enabling the measurement of thermal properties in ultrathin films and low-dimensional systems with sample masses in the nanogram range or even below[3,5,6]. The miniaturization of calorimetric cells (CCs), enabled by advances in microfabrication, has led to the development of membrane-based devices characterized by ultralow heat capacity addenda and excellent thermal coupling between the sample, heater, and sensors. As first noted in early studies of low-temperature calorimetry, the background signal—proportional to the total heat capacity of the system—can obscure the thermal signature of small samples if the addenda is not minimized[7]. Reducing the heat capacity of the calorimetric platform and matching it to that of the sample significantly enhances the effective SNR. Furthermore, the improved thermal contact in membrane-based chips nearly eliminates thermal lag and allows for much faster heating and cooling rates. Since the calorimetric signal scales with the heating rate, increasing the scan rate directly improves sensitivity. However, the scan rate must remain compatible with the intrinsic thermal diffusion timescales of both the sample and the calorimetric cell to ensure accurate measurements.

While existing nanocalorimetric techniques can characterize nanostructures with high sensitivity, they typically lack spatial (x-y) resolution and offer only depth information. In contrast, the proposed μs-PHnC technique overcomes this limitation by enabling local and selective characterization in the lateral (x-y) plane, allowing for precise thermal analysis of microscale features.

Three main nanocalorimetric methods have been implemented for chip calorimetry: AC calorimetry[8-11], relaxation calorimetry[1,5,12], and fast scanning calorimetry, which includes both custom microfabricated devices[3,13] and commercial implementations such as the Flash-DSC developed by Schick[14] and Mettler-Toledo. Flash-DSC, for instance, achieves high scan rates and can measure nanogram-scale samples, but its resolution is insufficient to capture subtle thermal events in ultrathin films and it remains limited by thermal diffusion length and platform design. In this work, we will focus on custom microfabricated devices used with fast scanning calorimetry, a technique often referred to as quasi-adiabatic nanocalorimetry (QAnC), since the microsecond-pulsed-heating nanocalorimetry technique presented in this paper is developed from QAnC principles.

AC-calorimetry: Building on the original concept introduced by Corbino in 1910, who proposed determining the specific heat of solids at high temperatures using alternating currents[8]. Sullivan and Seidel later established the theoretical foundation for steady-state AC calorimetry[9].This approach has since been adapted to nanoscale systems and implemented in various nanocalorimeters, enabling high-sensitivity measurements of the heat capacity of mesoscopic and low-dimensional materials. As a steady-state method, AC calorimetry permits the determination of heat capacity under quasi-isothermal conditions, either by combining it with temperature ramps or by probing its dependence on external variables such as magnetic field, electric field, time, or pressure. Notable examples include the attojoule-resolution calorimetry of mesoscopic superconducting loops[10]. Although the oscillatory nature of AC calorimetry enables the use of lock-in detection strategies that significantly enhance sensitivity, a key limitation of the technique is the requirement to operate below the characteristic thermal diffusion frequency of the device and its surroundings. This constraint ensures thermal equilibrium between the heater and sensor but inherently limits the maximum achievable heating rate and, consequently, the amplitude of the calorimetric signal. Strategies involving single-element heater-sensor configurations may improve internal thermalization times; however, they require the detection of higher-order harmonics, which are strongly attenuated and highly susceptible to noise. While this trade-off reduces the overall sensitivity, it enables the lateral sensing area to be minimized to the micrometre scale, making the technique suitable for spatially resolved studies in highly confined systems[11].

Relaxation calorimetry: A time-domain technique in which a sample is heated by a small, known amount, and the subsequent temperature relaxation back to equilibrium is monitored. The heat capacity is determined from the time constant of the exponential decay together with the thermal conductance of the system. If the thermal conductance between the cell and its surroundings is well-characterized, the heat capacity of the cell can be accurately determined. In this approach, the calorimetric signal remains relatively weak and, similar to the maximum achievable heating rates, is limited by the system's characteristic relaxation time. Nevertheless, the simplicity of this technique renders it highly scalable. Since its early implementation in 1972 by Bachmann et al.[1], through its adaptation in the first chip-based calorimeter[5] to its recent application in the measurement of 2D materials such as graphene with heat capacity resolutions down to $10^{-19}$ J/K[12], relaxation calorimetry has proven to be a versatile and powerful tool for thermal characterization at the micro- and nanoscale. However, although long thermal relaxation times are generally not ideal for fast or highly localized measurements, they become essential at very low temperatures. In this regime, the thermalization time between electrons and the phonon bath increases significantly, as discussed by Roukes[4]. For this reason, relaxation calorimetry remains the preferred technique for heat capacity measurements under such conditions at very low T. A prominent example is its implementation in the Physical Property Measurement System (PPMS) developed by Quantum Design, which enables precise calorimetric measurements across a wide temperature range.

Quasi-adiabatic nanocalorimetry (QAnC): Among the techniques implemented in chip calorimetry, QAnC stands out for offering the highest sensitivity per unit analysis area, with values down to 15 pJ $K^{-1}$ $mm^{-2}$. Originally developed by Leslie Allen and co-workers[3], it has since been implemented by many other research groups[3,13,15–18]. In this method, the calorimetric cell is rapidly heated several hundred degrees by Joule effect in few milliseconds using a single metallic element that acts as both heater and thermometer. The planar geometry and low thermal mass of the calorimetric cell enable high heating rates up to $10^5$ K/s, while maintaining moderate thermal homogeneity across the sensing area. Under high vacuum conditions, the calorimetric cell behaves adiabatically during the heating pulse, and the temperature evolution can be directly related to the input power. This configuration allows for the detection of subtle thermal events in ultrathin films with exceptional resolution[19–21].

However, the spatial resolution of QAnC is ultimately limited by the lateral thermal diffusion length, which defines the minimum effective analysis area sensed and, consequently, the minimal heat capacity addenda. When the analysis area approaches or falls below this limit, lateral heat spreading reduces temperature homogeneity and compromises the accuracy of the measurement. This limitation becomes particularly critical when attempting to measure individual samples that can only be fabricated or isolated with sufficient quality over restricted surface areas, as is often the case with certain two-dimensional (2D) materials.

Scaling down existing nanocalorimetric techniques often requires a trade-off between thermal homogeneity and measurement sensitivity. To address this, we present microsecond-pulse heating nanocalorimetry (µs-PHnC), a variant of quasi-adiabatic nanocalorimetry (QAnC) adapted to the microsecond timescale. The short duration of the pulses ensures that the temperature rise remains confined to the sensing region, preserving the quasi-adiabatic conditions required for direct heat capacity determination. This method retains the high sensitivity and temporal resolution of QAnC while incorporating several advantages of AC and relaxation calorimetry, namely, the ability to operate under quasi-isothermal conditions and to average over multiple pulses to reduce noise. Additionally, it enables the downscaling of the calorimetric cell to micrometre dimensions, but also shares some of the limitations of AC and relaxation techniques, such as reduced accuracy in measuring first-order transitions and limited capability for kinetic studies.

In this work, we detail the implementation of the technique, from the electronic architecture to the design and modelling of new nanocalorimeters with reduced sensing areas, addressing the technical considerations introduced by the shift in timescale. To validate the method, we selected CoO thin films, an antiferromagnet with a well-defined second-order Néel transition, ideally suited to the confined temperature profiles and high resolution of µs-Pulse Heating. Moreover, CoO can be deposited by thermal evaporation, making it highly compatible with chip-based calorimetry.

## 2. The us-pulsed nanocalorimetry

### 2.1. Methodology and device design

The microsecond-pulsed (µs-pulsed) nanocalorimetric method is a high-resolution thermal analysis technique derived from quasi-adiabatic scanning calorimetry. In this approach, the duration of the heating pulse is reduced to the microsecond range, producing a localized temperature increase of the calorimetric cell only a few Kelvin above the base temperature of the silicon frame. By applying pulse trains with controlled duty cycles, microseconds ON and milliseconds OFF, the system remains near thermal equilibrium, enabling multiple-pulse averaging to enhance signal-to-noise ratio and measurement precision. Figure 1 (a) shows a schematic of the train of pulses methodology. This configuration allows heat capacity to be evaluated under quasi-static conditions, as in AC or relaxation calorimetry, while retaining the high heating rates and sensitivity characteristic of QAnC.

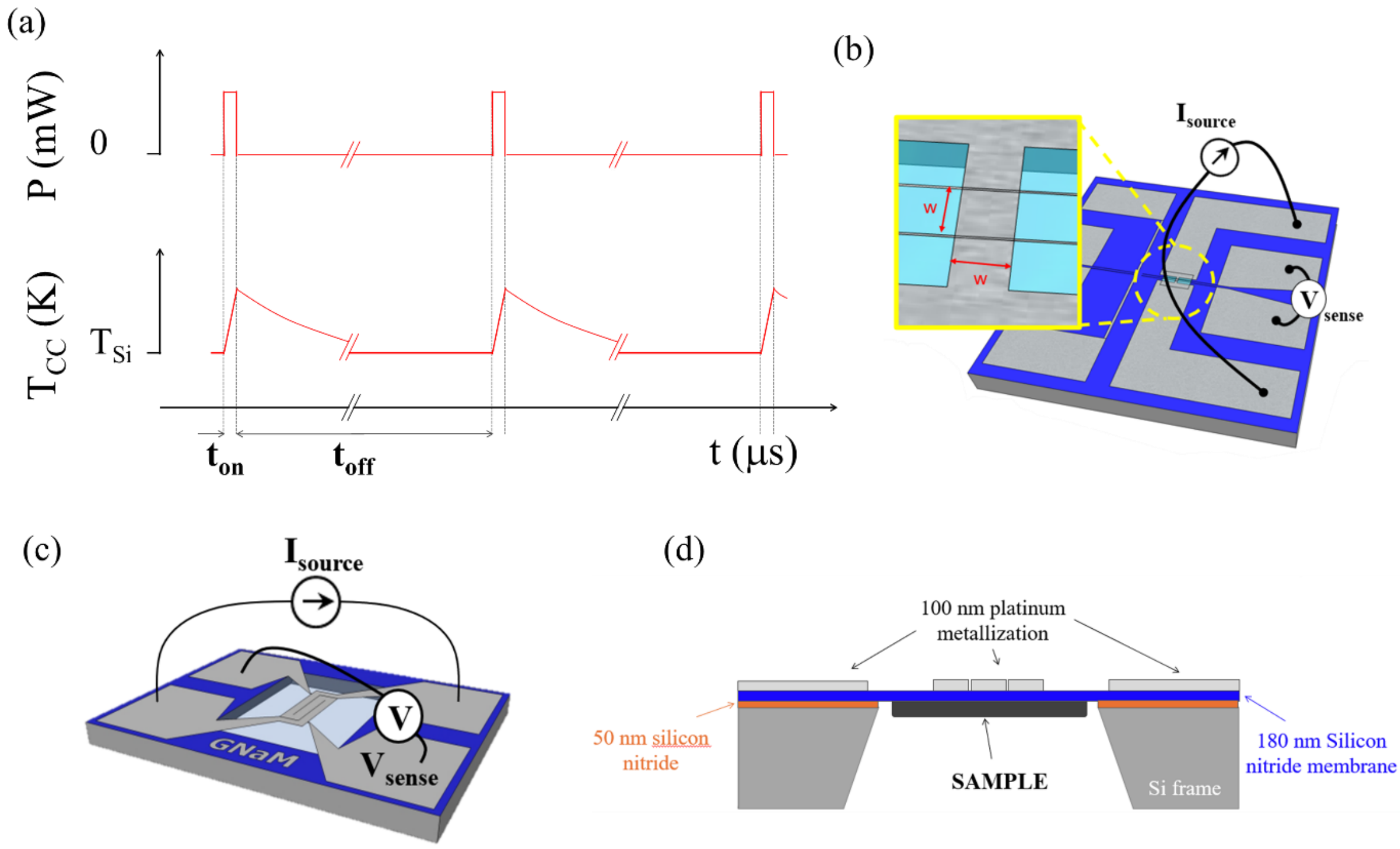

**Figure 1: Overview and working principle of the µs-pulsed nanocalorimeter.** (a) Train of current pulses applied to the nanocalorimeters. The duty cycle—defined as the ratio of pulse duration to total cycle time—is adjusted to ensure complete thermal relaxation of the calorimetric cell between pulses, maintaining the base temperature ($T_{Si}$) and enabling high-resolution averaging. (b) Strip-shaped heater/sensor with a square sensing area of side w (w = 30 µm, or 50 µm). Inset: magnified view of the strip-like sensor. (c) Meander-shaped heater/sensor with a sensing area of 1 mm². (d) Cross-sectional schematic of the calorimeter, showing the silicon nitride membrane supporting the calorimetric cell at the centre, including the metallic heater/sensor and the sample positioned underneath.

Figure 1 (b-d) presents different schematic views of the nanocalorimeters employed in this study. Each device consists of a freestanding silicon nitride ($Si_3N_4$) membrane, 180 nm thick, suspended over a silicon frame. At the centre of the membrane, a Ti/Pt bilayer (15/150 nm) defines the metallic element that functions both as a resistive heater and as a thermometer. Three distinct heater geometries were fabricated and tested: one meander-type design with a sensing area of 1 mm², and two strip-type designs with square sensing areas of 50 × 50 µm² and 30 × 30 µm², respectively. Samples, when present, are deposited on the backside of the silicon nitride membrane.

## 2.2. Principle of operation

Quantitative calorimetric measurements, irrespective of the specific architecture of the calorimetric cell, require the simultaneous determination of three key quantities: the time-dependent temperature of the cell, the electrical power input, and, under non-adiabatic conditions, the heat losses to the surroundings.

To operate the calorimeters, thermal excitation is achieved by injecting a controlled current $I(t)$ through the metallic element. The resulting Joule heating induces a rapid temperature increase in the calorimetric cell, including the underlying dielectric membrane. The metallic element is designed in a four-probe configuration, allowing for localized measurement of the voltage drop $V(t)$ at the calorimetric cell. The instantaneous power dissipated in the sensing area is calculated as $P_{in}(t) = V(t) \cdot I(t)$, and the time-dependent resistance is given by $R(t) = V(t)/I(t)$. These quantities, combined with a prior calibration of the resistance vs temperature relationship, enable accurate reconstruction of the temperature evolution $T(R(t))$ and subsequent heat capacity analysis. After calibration, the platinum metallization exhibits an electrical conductivity of approximately $6.6\times10^{6}$ S/m, with a temperature coefficient of resistance (TCR) of ≈ 0.0026 $K^{-1}$. At room temperature, the measured resistance is approximately ≈ 20 Ω for the meander-type design and ≈ 1.1 Ω for the strip-type sensors, with the sheet resistance being consistent across the different strip sizes. When injecting currents on the order of tens of milliamperes, the metallic element releases sufficient power to achieve heating rates on the order of $10^{5}$ K/s. Operating at such a high heating rate the nanocalorimeter behaves quasi-adiabatic.

Under high vacuum conditions, heat losses from the calorimetric cell are limited to conduction through the membrane and thermal radiation and operating at such a high heating rate the nanocalorimeter behaves quasi-adiabatic. For instance, a meander-type calorimeter requires approximately 80 mW to reach a heating rate of $10^{5}$ K/s, while the effective thermal conductance at room temperature is around 4 μW/K. Consequently, only after a temperature rise of ≈ 200 K, the power loss term $P_{losses}$ (4 μW/K x 200K) reaches ≈ 1% of the input power $P_{in}$. Therefore, for short scans, heat losses can be neglected, while for longer scans, they must be considered. Details on how to account for these losses are provided in the Supplementary Information. In quasi-adiabatic conditions, nearly all the input power contributes to the internal energy of the calorimetric cell, and the apparent heat capacity $Cp_{\beta}^{cc}(t)$ measured approaches the real heat capacity of the calorimetric cell $Cp_{real}^{cc}(t)$, since $\frac{P_{losses}(t)}{\beta(t)}$ becomes negligible:

$$Cp_{\beta}^{cc}(t) = \frac{P_{in}(t)}{\beta(t)} = Cp_{real}^{cc}(t) + \frac{P_{losses}(t)}{\beta(t)} \quad (1)$$

where $\beta(t) = dT(t)/dt$ the measured heating rate.

In this framework, the measurement of a sample requires two consecutive measurements of the same calorimetric cell: first with the empty CC (denoted with superscript 0) and subsequently with the sample evaporated onto it (denoted with superscript 1).

$$C_P{}^{sample}(T) \approx C_P{}_{\beta}^{cc,1}(T) - C_P{}_{\beta}^{cc,0}(T) \quad (2)$$

Analysis of the intrinsic noise in this measurement reveals that it is largely dominated by the numerical differentiation of digitized signals. Implementing true differential measurements significantly enhances the resolution.

The differential methodology can be implemented by connecting a pair of matched nanocalorimeters in series and performing the differentiation using a bridge of instrumentation amplifiers, as illustrated in Figure 2. The differential voltage drop between the two calorimetric cells, defined as $\Delta V = V_S - V_R$ where cell S hosts the sample (if any) and cell R serves as the reference, can be acquired with high amplification (typically ×10 to ×1000) using an instrumentation amplifier. The difference in apparent heat capacity between the two cells can then be recalculated by incorporating the differential voltage ΔV into the general expression (the complete derivation is provided in Supplementary Information)[16,22]:

$$\Delta C_P(T_S(t)) = \frac{I\,\Delta V}{\beta_S} - \frac{V_R}{\beta_S\,\beta_R}\frac{(d\Delta V/dt)_t}{(dR_S/dT_S)_t} + \frac{V_R\,I}{\beta_S}\left[1 - \frac{(dR_R/dT_R)_t}{(dR_S/dT_S)_t}\right] \quad (3)$$

Finally, in a complete experiment, the measurement of the sample's heat capacity must account for the difference between the differential heat capacities of the calorimetric cells before and after sample deposition, in order to subtract any pre-existing offset (similarly to the procedure followed in non-differential measurements):

$$C_P{}_{diff}^{sample}(T_S(t)) = \Delta C_P{}^1 - \Delta C_P{}^0 \quad (4)$$

While baseline and power losses corrections are discussed above, a heating-rate correction is not required here, since in the microsecond-pulse regime the relative delay between the temperature responses of the two cells is negligible.

## 2.3. Instrumentation and Electronics

The fast temporal dynamics of microsecond-pulsed nanocalorimetry necessitate a dedicated high-speed electronic setup. The system is built around a differential instrumentation amplifier bridge, where a pair of matched nanocalorimeters are connected in series with a precision load resistor. A schematic of the complete electronic chain is provided in Figure 2 (a). Signal generation and acquisition are performed using a PXIe-6124 data acquisition card (National Instruments), which offers 16-bit resolution and supports simultaneous analog input/output at sampling rates up to 4 MS/s within a ±10 V range. Synchronization and control of the pulse sequences are managed via LabVIEW software, which orchestrates the generation of current pulses by modulating the analog output channel (DAQ-AO). The pulse parameters - including amplitude, pulse width ($t_{on}$), and inter-pulse delay ($t_{off}$) -are fully programmable, enabling precise control of the duty cycle.

To ensure stable current delivery across varying load conditions, a Howland current source was implemented using an AD8519 operational amplifier. This voltage-to-current converter maintains a constant current through the bridge, independent of load variations, and can deliver up to 67 mA in the microsecond regime.

The different voltage signatures are pre-amplified using INA103 instrumentation amplifier, prior to the acquisition. The INA103, originally designed for low-noise acoustic applications, was selected for its fast-settling time (3 µs), ultra-low input noise ($1nV/\sqrt{Hz}$ at 1 kHz), and good common-mode rejection ratio for the amplification factors required in the experiment. The current flowing through each branch is monitored via the voltage drop across the load resistor ($V_I$), enabling accurate reconstruction of the power input and thermal response.

To conduct the experiments, the nanocalorimeters were mounted inside a custom-built liquid nitrogen immersion cryostat operating under high vacuum conditions (working pressures well below $10^{-6}$ mbar). The devices were mechanically secured to an aluminum mount holder using clamps, and thermal contact was enhanced with either Apiezon N grease or silver paste. Given the high sensitivity of the measurements to base temperature fluctuations ($\delta T_0$), the temperature of the holder was actively controlled. For this purpose, the mount was equipped with a resistive heater powered by a Keithley 2400 source meter and monitored using a cryogenic Pt100 thermometer connected to a Keithley 2700 multimeter. A LabVIEW®-based control program implementing a PID algorithm was used to regulate the temperature setpoint. The use of high-precision source meters (6.5-digit resolution) enabled excellent thermal stability, achieving peak-to-peak temperature fluctuations below 1 mK. This setup allowed for both isothermal and temperature-ramp experiments across the full operational range of 80 to 450 K, a range limited not by the µs-PHnC technique but by the cryostat used and the low-temperature response of the Pt heater/sensor.

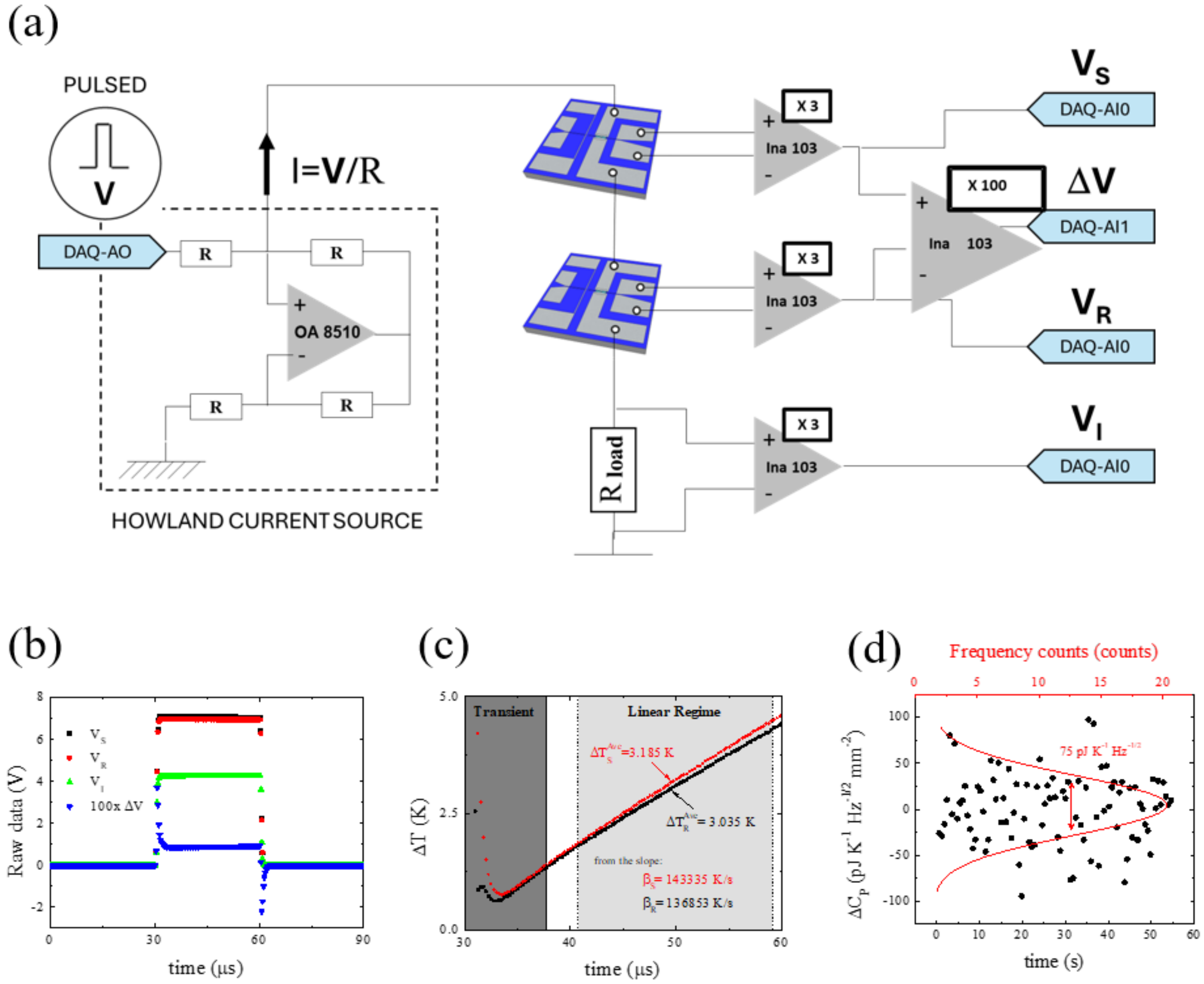

**Figure 2: Electronic Chain schematic:** (a) (On the left) The 'Howland' Current Source based on a Op. Amplifier 8510, (On the right) an amplification bridge based on INA103, permit to measure the voltage dropped in the calorimetric devices, the differential signal and the current passing through a load resistor. (b) Raw data voltages acquired for a single pulse. (c) Temperature evolution of a couple of nanocalorimeters, showing transient zone and linear zone used to perform the evaluation of the variables. (d) Fluctuations around the mean value of the difference in heat capacity between twin nanocalorimeters (meander type) measured fixed temperature (150 K), as a function of time. The dispersion of the data determines the uncertainty of the measurement.

Figure 2 (b) shows the raw voltage signals acquired during a single pulse experiment using a meander-type nanocalorimeter. The four recorded signals correspond to the sample ($V_S$), reference ($V_R$), current monitor ($V_I$), and the amplified differential signal ($100\times\Delta V$). The acquisition is configured to capture data both before and after the pulse, enabling baseline offset correction for each channel, an essential step for accurate averaging across multiple pulses. Figure 2 (c) presents the corresponding temperature evolution of the sample and reference calorimeters. Two distinct regions are highlighted: an initial transient zone (shaded dark grey), dominated by the response time of the INA103 amplifier (≈ 3 μs), and a subsequent linear regime (light grey), where the temperature increases linearly with time. From this linear region, the average temperature rise after 50 μs was determined to be 3.185 K for the sample and 3.035 K for the reference. The corresponding heating rates, extracted from the slope of the linear fits, were 143335 K/s and 136853 K/s, respectively. To assess the noise density of the pulsed heating method, heat capacity was measured over 50-second intervals. Figure 2 (d) displays the temporal fluctuations of the differential heat capacity signal ($\Delta Cp$) measured between two nominally identical nanocalorimeters at 150 K. Each black point corresponds to an individual measurement taken at a fixed acquisition rate. The dispersion of these values around zero is analyzed statistically, and the histogram of counts (top axis) is fitted by a Gaussian distribution (red curve). The standard deviation of this Gaussian represents the noise-equivalent fluctuation of $\Delta Cp$ within the acquisition bandwidth. To express the resolution in a bandwidth-independent form, we normalize this standard deviation by the square root of the measurement bandwidth, yielding units of $J\ K^{-1}\ Hz^{-1/2}\ mm^{-2}$. This allows comparing nanocalorimetry sensitivity across different acquisition times or instruments.

## 2.4. Thermal modelling and scalability of the calorimetric cell

Ideally, the temperature across the calorimetric cell should remain homogeneous during calorimetric experiments. This condition is partially met in rapid heating experiments, where power is locally dissipated in the sensing area of the heater, and lateral heat diffusion induces transient temperature gradients. In calorimetric cells with a $1\ mm^2$ sensing area, this diffusion remains moderate even on millisecond timescales, allowing quasi-adiabatic operation. However, with the implementation of the microsecond-pulsed nanocalorimetry (μs-PHnC) method, the temperature profile improves substantially.

Finite element modeling of devices with silicon nitride membranes of 180nm thick and 150nm of Pt metalization (details provided in the Supplementary Information) reveals that under 50 μs pulses at heating rates of 100 kK/s, approximately two-thirds of the sensing area exhibit temperature differences below 2.5 K. In contrast, for millisecond-scale pulses, temperature variations can reach up to 20 K across two-thirds of the calorimetric cell. Although the thermal inhomogeneity inherent to the meander-type design can be mitigated by integrating additional metallic spreaders within the sensing area, this approach introduces a significant increase in heat capacity addenda, thereby compromising the overall sensitivity of the calorimetric cell. [21]

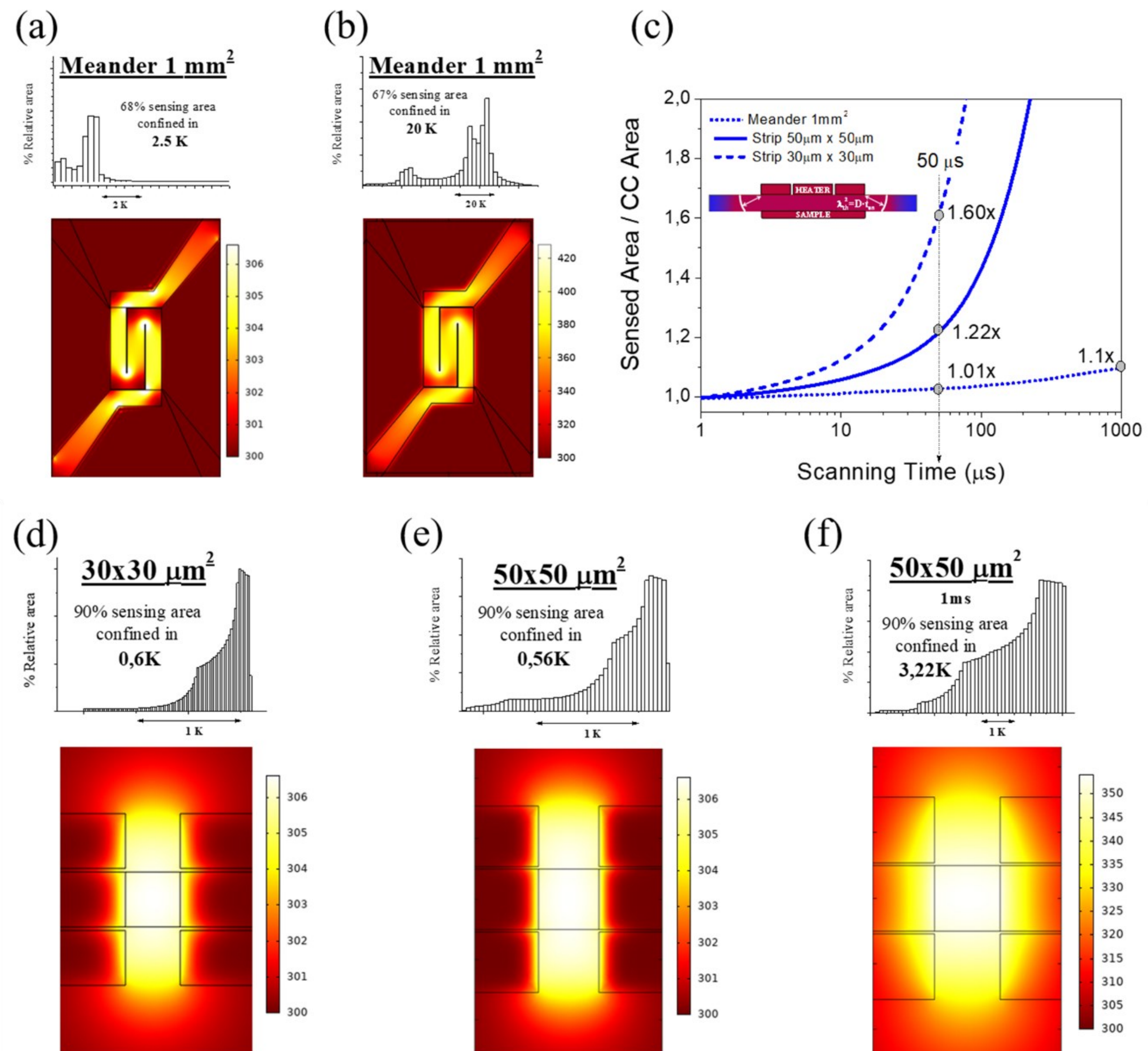

**Figure 3: Temperature maps of a meander like calorimeter heated at $10^5$K/s** at times of (a) 50 μs and (b) 1ms. The histograms show the 2D temperature dispersion in the calorimetric cell: showing that 2/3 of surface area is within 2.5 K after 50ms from the start of the pulse while the temperature inhomogeneity degrades to 20 K after 1ms. (c) Heat capacity of the same sample, 20 nm CoO, measured by QAnC (black squares), and μ-PHnC (red circles). (d-f) Show Simulated temperature maps for strip-type nanocalorimeters with sensing areas of 30 × 30 μm² and 50 × 50 μm². At 50 μs, 90% of the sensed area remains within a temperature range of 0.6 K and 0.56 K, respectively, indicating high thermal homogeneity. For the 50 × 50 μm² device at 1 ms, the temperature spread increases to 3.22 K, highlighting the impact of thermal diffusion over time on spatial uniformity.

Figure 3(a) and (b) display the simulated temperature maps and corresponding histograms for a meander-type calorimetric cell (1 mm²) at 50 μs and 1 ms, respectively.

To assess how spatial temperature degradation under fast heating affects heat capacity measurements and limits calorimeter miniaturization, we used finite element modelling to show that the sensed heat capacity increases with thermal diffusion length, which scales with the square root of membrane diffusivity and time. The simulations reveal deviations in the measured heat capacity, and thus in the effective sensed area or heat capacity addenda, from the intrinsic values of the calorimetric cell. Figure 3 (c) shows the normalized sensed area over time for different device geometries, based on comparisons between the heat capacity measured at a heating rate of 100 kK/s and the intrinsic heat capacity at the corresponding temperature, The inset schematically illustrates the thermal diffusion length λ, given by $\lambda^2 = D_{th} \cdot time$, where $D_{th}$ is the thermal diffusivity of the membrane. The analysis includes meander-type devices with 1 $mm^2$ sensing areas and strip-type nanocalorimeters with 50 × 50 $\mu m^2$ and 30 × 30 $\mu m^2$ sensing areas.

For meander-type devices with a ≈ 1 $mm^2$ sensing area, the sensed area increases by only ~1% after 50 μs (within the resolution limit) and reaches ≈ 10% after 1 ms. This moderate expansion follows a time dependence of $t^{1/2}$ and remains manageable even at millisecond timescales. In quasi-adiabatic nanocalorimetry (QAnC), this effect can be mitigated by using differential measurements and by confining the sample to the original calorimetric cell with a shadow mask. In contrast, for miniaturized strip-type devices with sensing areas of 50 × 50 $\mu m^2$ and 30 × 30 $\mu m^2$, the influence of thermal diffusion becomes significantly more pronounced. After 50 μs, the sensed area increases by approximately 22% and 60%, respectively. For these devices, the sensed area doubles the original calorimetric cell area after just 220 μs (50 × 50 $\mu m^2$) and 80 μs (30 × 30 $\mu m^2$), clearly demonstrating that long-duration pulses are unsuitable for small-scale designs. The rapid expansion of the sensed area severely limits sensitivity and imposes a fundamental constraint on the scalability of calorimetric cells and their applicability in QAnC. The μs-PHnC method, operating in the tens of microseconds regime, effectively minimizes this diffusion-driven expansion. By maintaining a reduced sensed area, it also limits the heat capacity addenda, a key factor in enhancing sensitivity. This makes μs-PHnC particularly well-suited for high-resolution calorimetry in ultrasmall samples.

Figure 3 (d-f) presents simulated temperature profiles for strip-type devices with sensing areas of 30 × 30 $\mu m^2$ and 50 × 50 $\mu m^2$. At 50 microseconds, the heated region closely matches the heater footprint, which represents the ideal scenario. However, a narrow region surrounding the heater also becomes heated due to thermal diffusion, extending a few micrometres beyond the active area. This effect is more noticeable in the smaller device and becomes increasingly significant at longer pulse durations, as shown in Figure 3 (c).

The improvement in temperature homogeneity is evident. For the 50 × 50 μm² calorimeter, 90 percent of the sensed area falls within a temperature range of 0.56 K at 50 μs, as shown in the histogram. As expected, increasing the pulse duration or reducing the device size leads to greater temperature dispersion. For example, at 50 μs, the 30 × 30 μm² device shows a temperature spread of 0.6 K, while the 50 × 50 μm² device reaches 3.22 kelvin at durations of 1ms. These results highlight the trade-off between temporal resolution and spatial uniformity in miniaturized calorimetric designs.

## 2.5. Experimental validation of transient response

The process of scaling down is not entirely free from challenges. In the initial implementations, we used devices with long strip heaters, several millimetres in length, which exhibited inductive effects that hindered accurate temperature sensing during the early stages of the pulse. Geometrical analysis of the heater's self-inductance showed that it is proportional to the strip length and to the logarithm of the inverse of its transverse dimensions, namely width and thickness. This means that longer strips, or those narrower or thinner, exhibit stronger inductive behaviour.

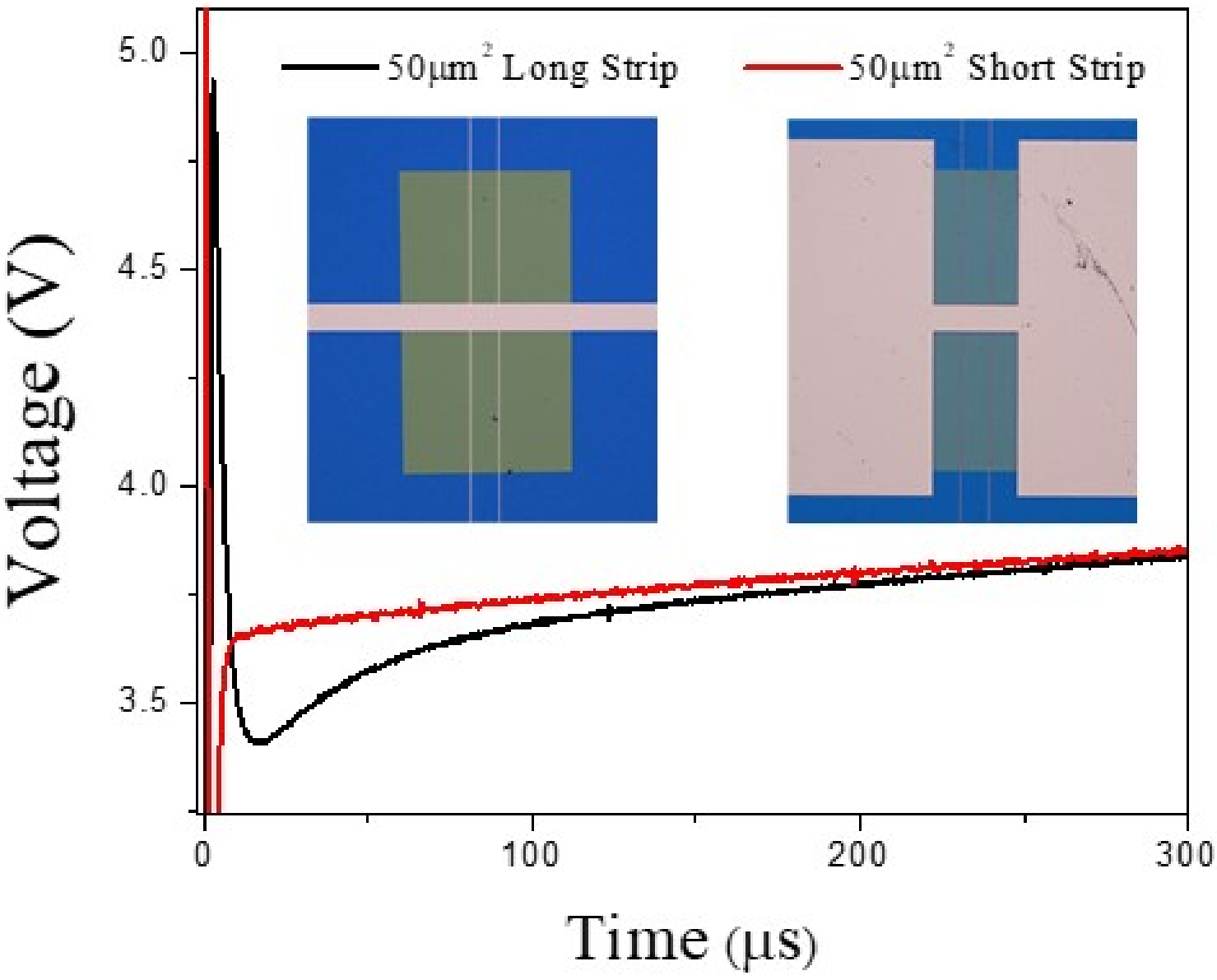

**Figure 4:** Minimizing Inductive Artifacts in Miniaturized Nanocalorimeter Designs. Voltage response during a single pulse for devices of different sizes and configurations, highlighting the improved signal linearity in the optimized design. (Red curve): Original 50 × 50 μm² strip-type heater layout showing inductive distortion due to extended lead length. (Black curve): Optimized design with shortened leads, effectively suppressing inductive effects and enhancing signal linearity.

Since the strip width defines the calorimetric cell size and is fixed by design, we addressed this issue by increasing the metal thickness and, more effectively, by reducing the strip length, guided by the thermal diffusion constraints. Figure 4 shows the voltage response measured in a four-point configuration for two 50 × 50 µm strip-type calorimeters: one with a long strip several millimetres in length, and another with a short strip of 150 µm. The long-strip design exhibits an inductive transient lasting 150 µs, whereas in the short-strip design this transient is reduced to less than 10 µs, enabling accurate temperature sensing from the onset of the pulse.

## 3. Results and discussion

To demonstrate the capabilities of the microsecond-pulsed heating nanocalorimetry (µs-PHnC) technique and its scalability to reduced sensing areas, we fabricated three nanocalorimetric devices: one with a meander-type geometry (1 $mm^2$) and two with strip-type geometries featuring calorimetric cell sizes of 30 × 30 $µm^2$ and 50 × 50 $µm^2$, respectively. Thin films of cobalt monoxide (CoO) were deposited onto the backside of the calorimetric membranes using electron beam evaporation[23]. The antiferromagnetic transition of CoO was selected as a benchmark due to its continuous nature and absence of latent heat, making it an ideal candidate for evaluating the resolution of the technique.

Prior to sample deposition, as in quasi-adiabatic nanocalorimetry (QAnC), the baseline heat capacity difference between the calorimetric cells was measured using both non-differential and differential configurations. Heat capacity was recorded at 0.5 K intervals from 150 K to 380 K, averaging 1 second per temperature point. One of the main limitations of µs-PHnC compared to QAnC is the longer acquisition time required to achieve high-resolution measurements. This extended duration increases the susceptibility of the calorimetric cell to contamination from residual water or other adsorbates. To mitigate this, a preconditioning step was implemented before each measurement, consisting of either a rapid temperature ramp to 400 K or a flash heating cycle to 600 K. These procedures effectively desorb unwanted species from the calorimetric cell. When this protocol is followed, the measurements exhibit excellent reproducibility.

A key advantage of the µs-PHnC technique is that it eliminates the need for a shadow mask to confine the sample. The effective analysis area is inherently defined by the heater geometry and the thermal diffusion length corresponding to the pulse duration. In this study, a 20 nm CoO film was deposited on the meander-type device with the aid of a shadow mask that limits the sample to the active area of the calorimetric cell, while 95 nm and 70 nm CoO films were deposited shadowless on the 30 × 30 $µm^2$ and 50 × 50 $µm^2$ strip-type devices, respectively. Film thicknesses were estimated using a QCM and calibrated via profilometry, with an associated uncertainty of approximately 10%. Figure 5 (a-c) presents the differential heat capacity curves before and after CoO deposition for each of the calorimetric cell designs. The data clearly illustrate the thermal response of the system and the effectiveness of µs-PHnC in capturing the heat capacity signature of the CoO thin films across different device geometries.

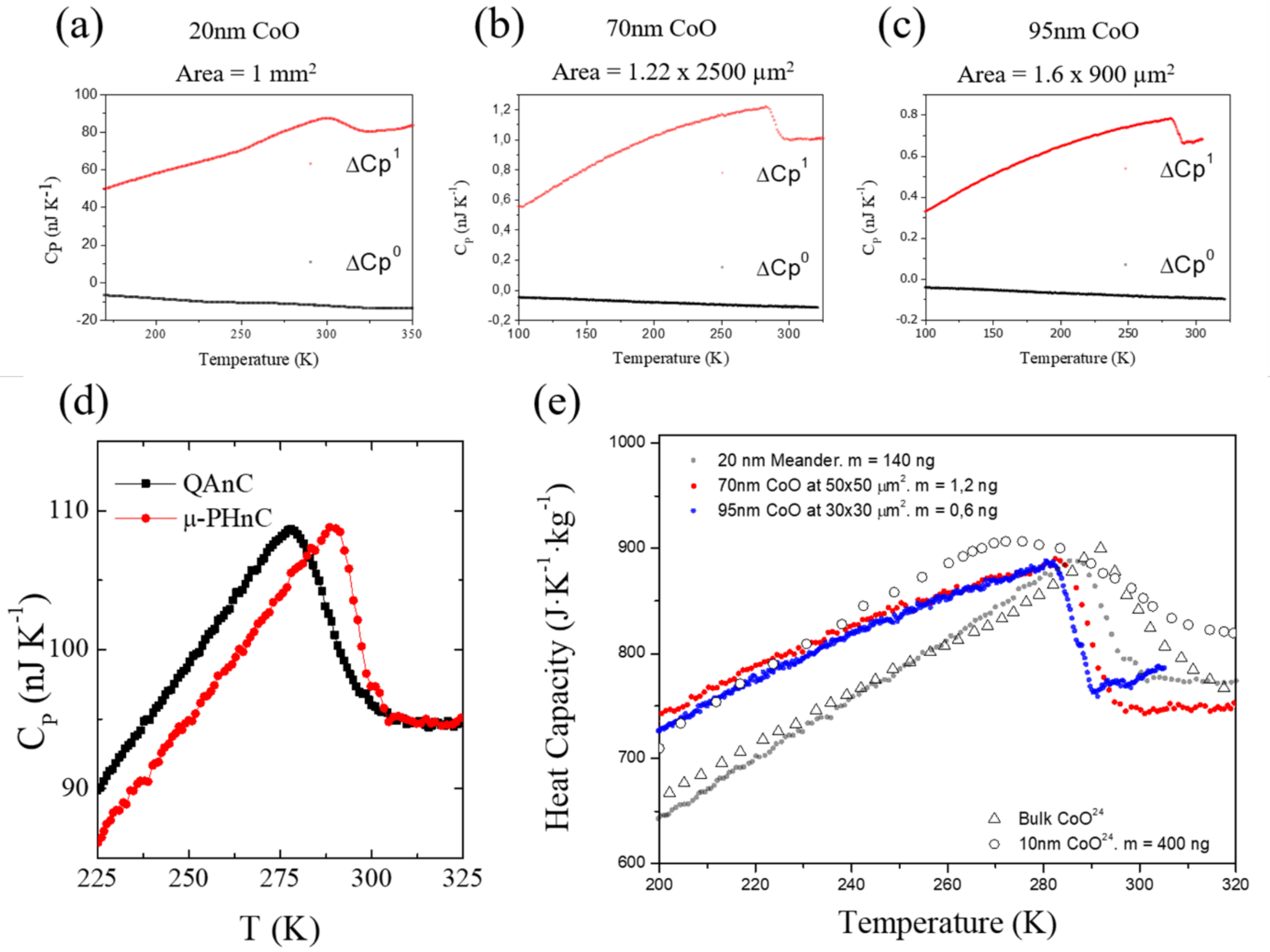

**Figure 5.** Differential heat capacity (ΔCp) measurements of CoO thin films with varying sensing areas and nominal thicknesses. The three panels correspond to: (left) a 20 nm CoO film measured on a meander-type calorimeter (1 mm² sensing area), (middle) a 70 nm CoO film on a 50 × 50 µm² strip-type device, and (right) a 95nm CoO film on a 30 × 30 µm² strip-type device. In each case, $\Delta Cp^0$ (black) represents the baseline differential signal prior to sample deposition, and $\Delta Cp^1$ (red) corresponds to the signal after sample deposition. (d) Heat capacity of the same sample in meander device, 20 nm CoO, measured by QAnC (black squares), and µ-PHnC (red circles). (e) Specific heat capacity of CoO thin films measured using µs-PHnC across different calorimetric cell geometries.

The improvement in temperature uniformity achieved with microsecond-pulsed heating nanocalorimetry (µs-PHnC), shown in Figure 3, is experimentally validated by analyzing the width of the calorimetric anomaly associated with the Néel transition in CoO thin films. For samples fabricated via physical vapor deposition (PVD), the grain size is expected to lie within the nanometre regime [23]. In CoO, this antiferromagnetic-to-paramagnetic transition appears as a lambda-type anomaly in the heat capacity curve. While this transition is sharp in bulk materials, thin films exhibit a broadened transition due to finite-size effects, particularly when the grain size approaches the correlation length of the superexchange interaction. Figure 5 (d) presents a comparative analysis of the heat capacity signal for a CoO film with a nominal thickness of ~20 nm (as determined by quartz crystal microbalance, QCM), measured using both quasi-adiabatic nanocalorimetry and μs-PHnC. In the quasi-adiabatic regime, the transition spans approximately 20 K, consistent with the thermal gradients predicted for millisecond-scale heating. In contrast, μs-PHnC measurements reveal a significantly narrower transition (≈ 5 K), which lies above the modelled temperature dispersion, thereby offering a more faithful representation of the intrinsic rounding induced by the nanocrystalline structure.

The subtraction of the differential heat capacity values before and after sample deposition yields the net heat capacity of the CoO thin film. To calculate the specific heat, we consider the effective analysis areas obtained from finite element modelling, the nominal thicknesses estimated via QCM calibration, and an assumed density of 6450 $kg \cdot m^{-3}$, corresponding to the bulk value for CoO. It is worth noting, however, that thin films deposited by physical vapor deposition (PVD) may exhibit slightly lower densities due to microstructural effects. Figure 5 (e) presents the calculated specific heat values for each device configuration. Red circles correspond to a 70 nm CoO film measured on a 50 × 50 $\mu m^2$ strip-type device; blue circles represent a 95 nm CoO film on a 30 × 30 $\mu m^2$ striptype device; and black circles correspond to 20 nm CoO films measured on meander-type devices. For comparison, literature data from Tang et al.[24] are included: open triangles for 10 nm CoO and open circles for bulk CoO. The sharp lambda-type anomaly near 290 K reflects the Néel transition, with the μs-PHnC technique capturing the transition with high resolution across all device configurations. The corresponding sample masses for each measurement are indicated in the legend. It is worth to highlight that strip-type μs-PHnC provides a ~ $10^3$- fold reduction in sample mass relative to previous study[24], while maintaining excellent signal-to-noise. The technique's high temporal resolution and spatially confined heating minimize instrumental broadening, yielding a sharply defined magnetic transition. The same attributes can be further exploited to resolve with unprecedented resolution the subtle transition rounding commonly attributed to finite-size effects when downscaling to thicknesses of only a few nm. Crucially, μs-PHnC resolves phase transitions in sub-nanogram specimens, paving the way for measurements on single-monolayer 2D materials and other ultra-small solids.

## 4. Conclusions

We have developed and validated a novel microsecond-pulsed heating nanocalorimetry (µs-PHnC) technique that enables high-resolution, quasi-isothermal heat capacity measurements in ultrathin films and low-dimensional materials, allowing thermal characterization in nanosystems that was previously beyond reach. This method builds upon the principles of quasi-adiabatic nanocalorimetry (QAnC) while overcoming its limitations in spatial resolution and scalability. By reducing the pulse duration to the microsecond regime (typically 50 µs), µs-PHnC minimizes lateral thermal diffusion, preserving temperature homogeneity and enabling accurate measurements even in calorimetric cells with sensing areas as small as $30 \times 30$ µm².
The technique was implemented using custom-fabricated nanocalorimeters with meander- and strip-type geometries, and its performance was benchmarked using the antiferromagnetic transition in CoO thin films. The results demonstrate that µs-PHnC captures the lambda-type anomaly with significantly improved sharpness compared to QAnC, reducing the transition width from ≈ 20 K to ≈ 5 K. Finite element modelling confirms that the sensed area remains confined to the heater footprint during short pulses, which is critical for maintaining sensitivity and minimizing heat capacity addenda. Consequently, there is little benefit in reducing the strip-type calorimeter size below ≈ $30 \times 30$ µm², since the thermal diffusion length linked to pulse duration in that range and prevents further reduction of the effective sensing area.
The µs-PHnC method also offers practical advantages: it eliminates the need for shadow masks, simplifies sample deposition, and supports integration with external stimuli under quasi-static conditions. Because the system remains near thermal equilibrium between pulses, it is ideally suited for coupling with slowly varying external fields—such as magnetic, electric, or optical stimuli—enabling the study of field-dependent thermodynamic responses with high temporal and thermal resolution.
The differential measurement scheme, combined with high-speed electronics and robust thermal modelling, ensures excellent reproducibility and noise performance, with a noise density as low as 75 $\mathrm{pJ \cdot K^{-1} \cdot \sqrt{Hz} \cdot mm^{-2}}$. The newly designed calorimeters further expand the capabilities of nanocalorimetry by allowing heat capacity measurements of materials with extremely small surface areas and heat capacity addenda below $10^{-9}$ J/K. This advancement paves the way for high-resolution heat capacity measurements of materials that were previously inaccessible. While traditional high-resolution Cp measurements were limited to thin-film materials, this technique now extends the capability to nanostructured and 2D materials with significantly smaller analysis areas. This represents a critical step forward in the study of thermodynamic properties of low-dimensional and nanostructured systems, broadening the scope of calorimetric applications in material science.

## Conflict of Interests

The authors declare no conflict of interests.

## Author contributions

H.G. and M.M. performed the experiments, analyzed the data, carried out the simulations. A.L., H.G. and M.M prepared the final version of the manuscript. S.P. and E.M. fabricated the CoO samples. L.A. fabricated the nanocalorimetric devices. O.B. and A.L. conceived the idea of the work. J.R. contributed to the scientific discussions. All authors contributed to the revision and approved the manuscript.

## Acknowledgements

The authors the financial support from: Project TED2021-129612B-C22 and TED2021-131363B/I00 funded by the European Union NextGenerationEU / PRTR (grant Nº CNS2022-135230), Projects PID2020-117409RB-I00 and PID2023-152783OB-I00 funded by MCIN/AEI /10.13039/501100011033) and grants 2021SGR-00644, 2021SGR-00497 and 2021SGR-00651 funded by AGAUR (Generalitat de Catalunya). HGT acknowledges support from the Spanish Ministry of Science and Innovation through the FPI fellowship PRE2021-097883. SP thanks DOCFAM-PLUS within HORIZON-MSCA-2021-COFUND-01 (grant Nº 101081337). The authors thank ICTS-SB-CSIC, which is part of the ICTs Micronanofabs node supported by the Spanish Ministry of Science, Innovation and Universities (MICIU). E. M. is a Serra Húnter fellow. JRV and ALF secured funding to this work. The Catalan Institute of Nanoscience and Nanotechnology (ICN2) is funded by the CERCA program/Generalitat de Catalunya and is supported by the Severo Ochoa Centres of Excellence Programme, Grant CEX2021-001214-S, funded by MCIU/AEI/10.13039.501100011033.

HGT and MMR contributed equally to this work.

# Supplementary Information

**Microsecond-Pulsed Nanocalorimetry: A Scalable Approach for Ultrasensitive Heat Capacity Measurements**

*Hugo Gómez-Torres[1], Manel Molina-Ruiz[2], Simone Privitera[2], Enric Menéndez[2], Llibertat Abad[3], Jordi Sort[1,4], Olivier Bourgeois[5], Javier Rodriguez-Viejo[1,2], Aitor Lopeandia[1,2*]*

[1] Catalan Institute of Nanoscience and Nanotechnology (ICN2), CSIC and BIST, E-08193 Bellaterra, Spain.
[2] Departament de Física, Universitat Autònoma de Barcelona, E-08193 Bellaterra, Spain.
[3] Institut de Microelectrònica de Barcelona (IMB-CNM-CSIC). Campus de la UAB, E-08193, Cerdanyola del Vallès (Barcelona), Spain.
[4] Institució Catalana de Recerca i Estudis Avançats (ICREA), Pg. Lluís Companys 23, E-08010 Barcelona, Spain.
[5] Institut Neel, CNRS et Université Joseph Fourier, 25 avenue des Martyrs, F-38042 Grenoble Cedex 9, France

*E-mail: aitor.lopeandia@uab.cat

## S1.- Finite Element Modeling of the Nanocalorimeters.

Finite element method (FEM) simulations are performed using COMSOL Multiphysics® using PDE modules to combine "heat transfers in solids' with "electrical conductive media". The thin film nature of the nanocalorimetric device membrane guaranties that vertical temperature gradients can be neglected permitting to mathematically reduce the problem to a 2D configuration, considering a constant $d_z$ thickness along the structure. The simplified two-dimensional model should consider the equivalent physical properties of the stack of materials for each physical domain, referenced to apparent thickness.

Under this approximation heat transfer equation can be rewritten as,

$$d_z \rho C_p \frac{\partial T}{\partial t} - d_z \nabla k \nabla T = d_z Q_e + q_o$$

where $\rho C_p$ is the density times the heat capacity, $T$ is the temperature, $k$ is the thermal conductivity, $Q_e$ is the electromagnetic Joule heating ($Q_e = J \cdot E$) and $q_0$ the out of plane heat radiative component,

$$q_0 = \epsilon_u \sigma \left(T_{ext,u}^4 - T^4\right) + \epsilon_d \sigma \left(T_{ext,d}^4 - T^4\right)$$

being $\epsilon_u$ and $\epsilon_d$, the up and down emissivity of the surface, $T_{ext,u}$ and $T_{ext,d}$ the external reference temperatures in both sides of the surface and $\sigma$ is the Stephan Boltzmann constant.

Convective heat losses components are not considered since nanocalorimeters normally operate at high vacuum pressures and thus can be neglected.

Under 2D approximation Maxwell equations can also be implemented considering the $d_z$ dimension on the evaluation of the extensive values of the electrical conductance.

For the different designs, the planar geometries of the nanocalorimeters under study were defined using the CAD tools available in COMSOL Multiphysics®. As an example, the resulting layouts and mesh for the strip heater with sensing area of 50µm x 50µm are shown in Figure S1. For each domain, the effective physical properties of the stacked materials were considered. In particular, the thermal conductivity $k$ for each region was calculated as described below:

$$k = \sum_{i=0}^{n} k_i \cdot \left(\frac{d_i}{d_z}\right)$$

where $k_i$ is the thermal conductivity of the $i$-th material layer, $d_i$ is its actual thickness, and $d_z$ is the apparent total thickness of the region in the model. This approach ensures that the contribution of each material is properly weighted according to its relative thickness within the stack.

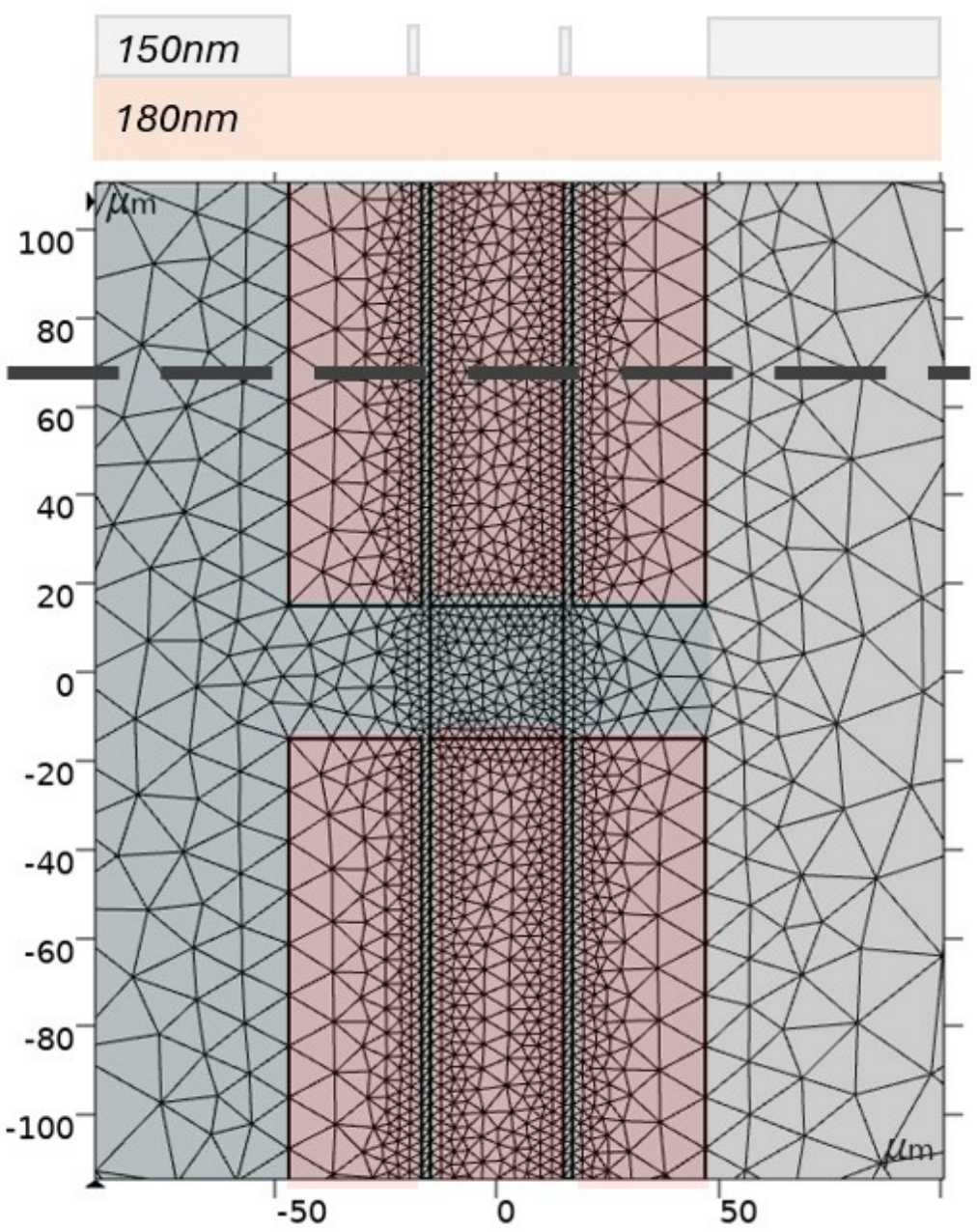

**Figure S1. Layout and mesh of a strip-type calorimetric cell.** Representative layout and finite element mesh of a calorimetric device with a 50 µm × 50 µm sensing area. The cross-sectional view highlights two distinct regions composed of stacked platinum and silicon nitride layers.

When defining the heat capacity of the regions, we considered that the density is also defined common to the region and material, and should be weighed as follows:

$$\rho C_p = \sum_{i=0}^{n} \rho_i C_{p,i} \cdot \left(\frac{d_i}{d_z}\right)$$

Material properties and their temperature dependence were obtained from literature and from previous experiments conducted in our laboratory. The thermal conductivity of the silicon nitride layer, including its temperature dependence, was determined from measurements using the Völklein method [1]. The thermal conductivity of the platinum layer was derived from the electrical resistivity of the heater strip, applying the Wiedemann–Franz law.

Figure S2 shows the electrical conductivity values used in the modelling. These values were obtained as averages from the resistivity extracted through resistance–temperature calibration of four calorimetric device designs (a meander-type and three strip geometries), all featuring 150 nm-thick platinum metallization. The resulting conductivity values enabled accurate reproduction of the experimental resistance across the different heater geometries in the simulations.

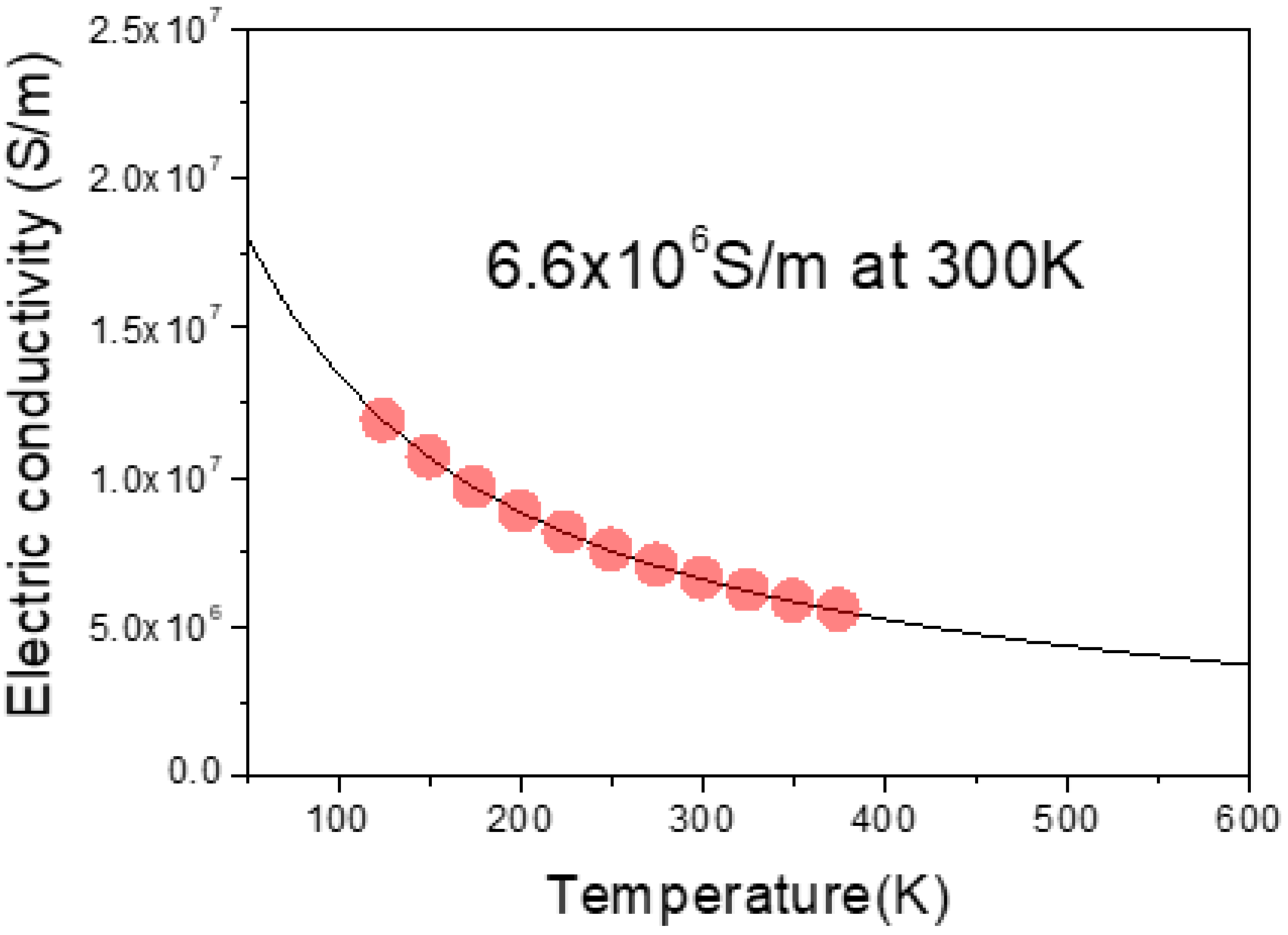

**Figure S2. Electrical conductivity derived from heater resistance calibration:** Electrical conductivity was extracted from the temperature-dependent calibration curves of the heater resistance, taking into account the geometrical parameters. For devices with a 150 nm-thick platinum metallization, the extracted conductivity values accurately reproduce the experimental resistance across different heater geometries in the simulations.

To determine the heat capacity values, we employed the Debye model, using as an upper bound the Dulong–Petit limit: 1050 $J \cdot K^{-1} \cdot kg^{-1}$ for silicon nitride and 135 $J \cdot K^{-1} \cdot kg^{-1}$ for platinum. The temperature dependence was introduced by adjusting the Debye temperature, set at 225 K for the platinum thin film and 550 K for the silicon nitride membrane. These adjusted values are lower than those reported for bulk materials, which is consistent with the expected reduction in Debye temperature due to finite-size effects in the thin films. Since the variation of the mass density is very small in the temperature range of the experiments, we considered constant values of 3100 $Kg \cdot m^{-3}$ and 21400 $Kg \cdot m^{-3}$, for silicon nitride and platinum, respectively.
Among the various parameters, the actual thickness $d_i$ of the constituent layers carries the greatest uncertainty and was therefore treated as an adjustable parameter in the modelling. Using the pulsed heating technique, we experimentally measured the heat capacity of the empty calorimetric cell, and applying the same procedure to the simulated data we can compare the results. In this context, increasing the thickness of either the silicon nitride layer $d_{SiNx}$ or the platinum layer $d_{Pt}$, significantly impacts the total heat capacity of the system.
Fortunately, the transient temperature response to a current pulse allows us to decouple these two contributions and refine the parameter estimation. Given that platinum has a thermal conductivity approximately 25 times higher than that of silicon nitride, even small variations in the estimated thickness noticeably affect the thermal transient. A good agreement between experiment and simulation was achieved by assuming a platinum thickness consistent with the nominal range (150 nm) and a silicon nitride thickness of approximately 173 nm, close to the nominal value of 180nm. Figure S3 illustrates both the agreement between the simulated and measured heat capacity of the empty calorimetric cell, and the fidelity of the model in reproducing the transient thermal response under a current pulse.
Although radiative losses can be considered negligible in microsecond-scale pulsed experiments—since the temperature increase is only a few kelvin above the base temperature—we included emissivity values obtained from previous nanocalorimetric measurements. The upper emissivity $\epsilon_u$ was set to approximately 0.1 for domains capped with platinum, and around 0.54 for domains exposing bare silicon nitride. At the experimental temperatures, the dominant infrared wavelengths have a penetration depth greater than the characteristic thickness of the silicon nitride membranes. As a result, the effective emissivity is expected to be lower than the bulk value (≈ 0.9), where deeper atomic layers also contribute to radiation. For the emissivity on the dorsal side $\epsilon_d$, we also considered a value of 0.54 for the bare silicon nitride membrane. When the membrane is covered with platinum on the upper side, we accounted for the additional contribution of the platinum layer—due to the partial transparency of the silicon nitride—resulting in a total effective emissivity of approximately 0.64.

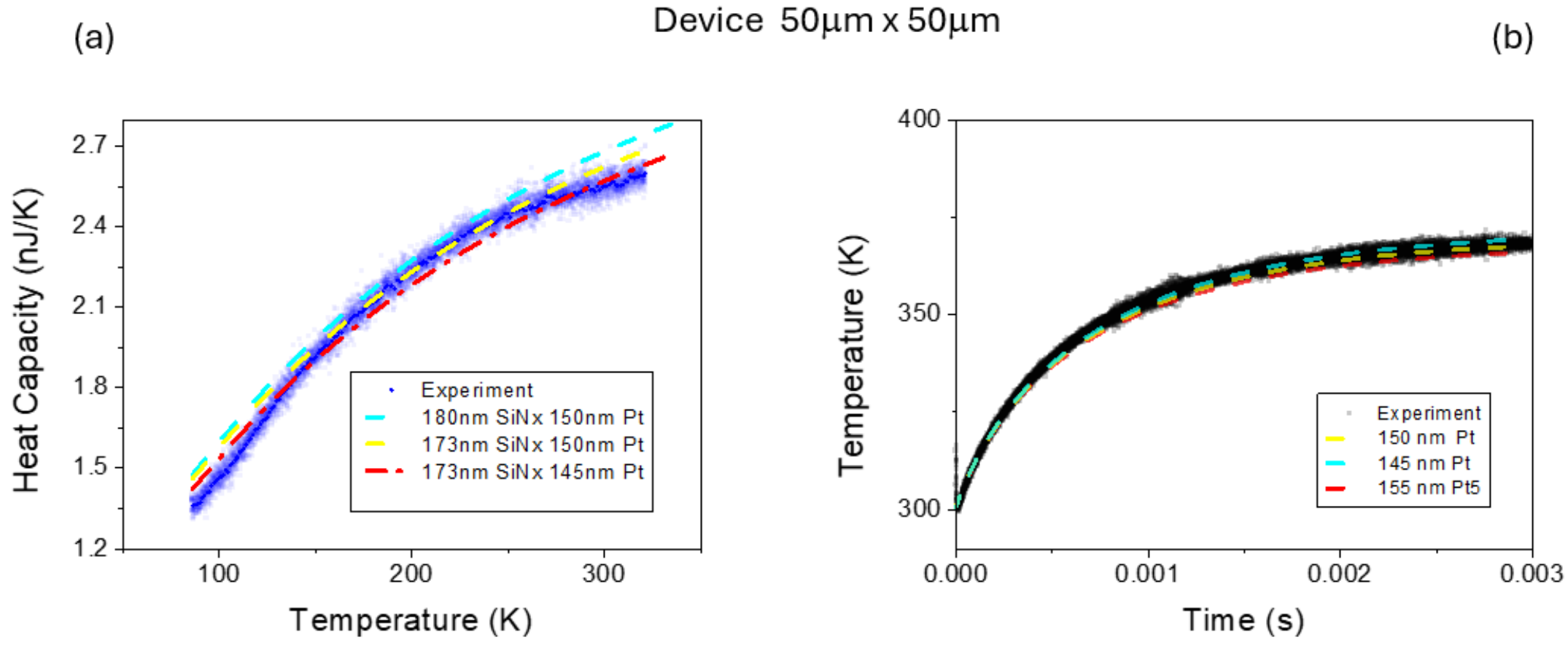

**Figure S3. Adjustment of material parameters via pulsed heating.** (a) Heat capacity measurement of an empty calorimetric cell (50 µm × 50 µm) using a pulsed heating technique, compared with finite element method (FEM) simulations incorporating adjusted material parameters. (d) The temperature evolution under a constant 17.5 mA current pulse is shown, with FEM simulations considering varying platinum metallization thicknesses to assess their influence on thermal response.

FEM simulations allow us to determine temperature profiles at specific time points. In addition to this, we used FEM simulations to reproduce the experimental response observed when measuring a thin film of cobalt monoxide (CoO). Cobalt monoxide exhibits a characteristic lambda-shaped anomaly in its heat capacity associated with the antiferromagnetic transition near 290 K. Given that antiferromagnetic superexchange interactions are typically sensitive to finite-size effects, we considered heat capacity data reported in the literature for nanometric CoO samples [2].

Figure S4 shows the heat capacity curves for bulk CoO and for a sample with 10 nm grain size, both of which were implemented as material models in COMSOL Multiphysics ®. For the in-plane thermal conductivity, we used a value of 0.9 $W \cdot m^{-1} \cdot K^{-1}$, measured on a 50 nm-thick film using the Volklein technique. The density was set to the tabulated bulk value of 6450 $kg \cdot m^{-3}$ at room temperature. However, since the samples are polycrystalline thin films deposited by electron beam evaporation, it is expected that the actual density may be slightly lower.

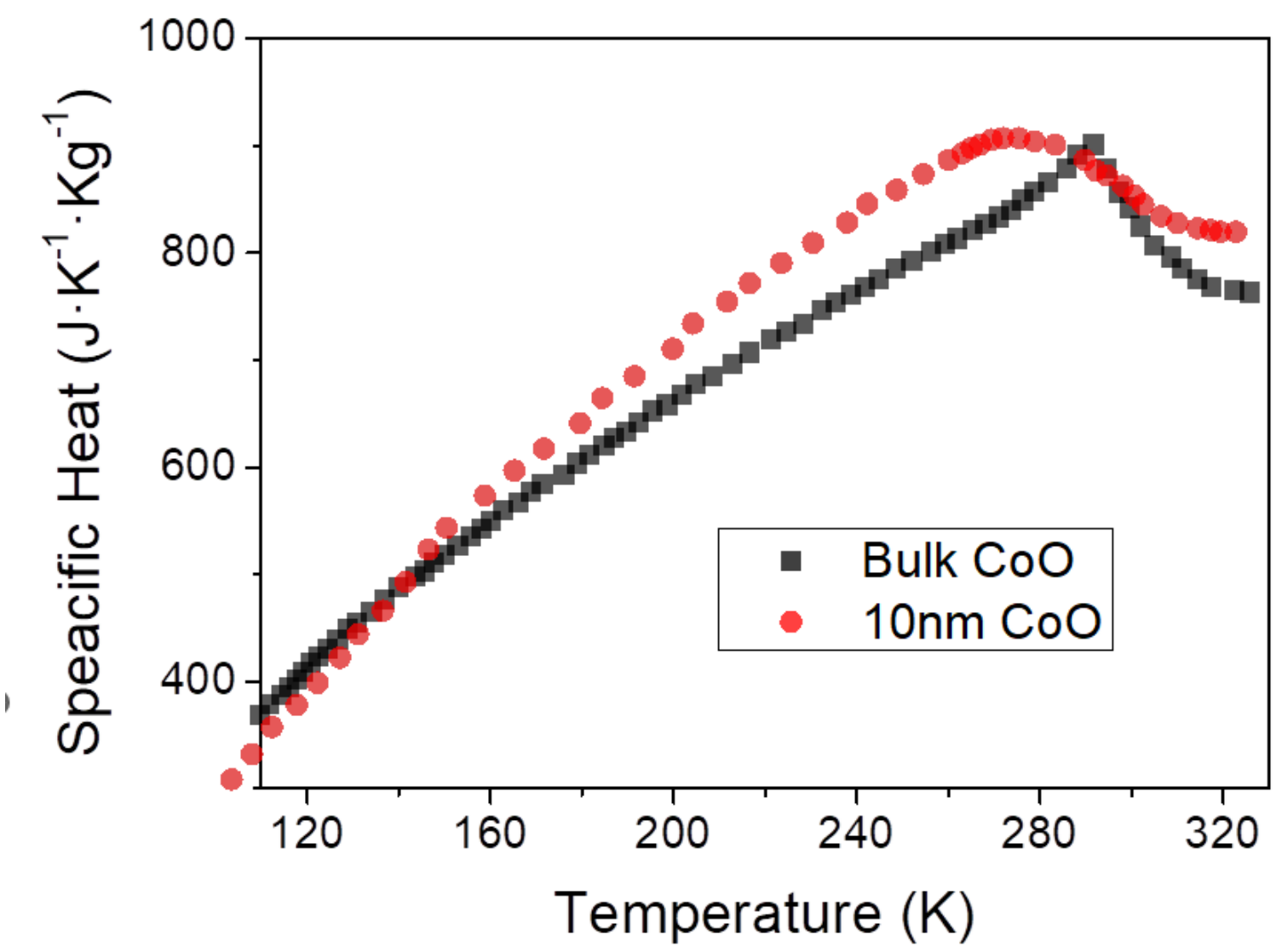

**Figure S5. Specific heat capacity of CoO in bulk and nanostructured form.** Temperature-dependent specific heat capacity of cobalt monoxide in bulk form and as thin films with 10 nm grain size. The data were implemented in COMSOL Multiphysics ® to model the thermal response of nanocalorimetric measurements. Figure adapted from Tang et al.[2]

## S2.- Heat capacity analysis.

The heat capacity analysis in the nanocalorimetric technique based on microsecond pulse heating builds upon the quasi-adiabatic approach, leveraging the possibility of linearizing the temporal evolution of the relevant variables. For completeness, we include in the supplementary material a detailed analysis of the quasi-adiabatic method, which has been extensively described in the literature[1,3].

In membrane-based nanocalorimetric devices, the calorimetric cell incorporates a thin metallic layer that functions simultaneously as a resistive heater and a temperature sensor. This configuration enables the delivery of thermal power via Joule heating and allows for the measurement of the average temperature of the cell during thermal scans.

Throughout the heat capacity analysis, it will be assumed that temperature gradients within the calorimetric cell are moderate, thereby justifying the use of the average temperature as a representative value for the system, nevertheless, this assumption has limitations. Due to the reduced out-of-plane dimensions of both the calorimetric cell and the sample, the in-plane thermal conductance is inherently limited. As previously evaluated[4,5], the spatial temperature distribution is highly dependent on the applied heating rate and the presence of additional thermal diffusion layers within the calorimetric structure. When the heating rate is low compared to the characteristic lateral cooling rates, significant inhomogeneities can arise in the two-dimensional temperature distribution, typically resulting in parabolic temperature profiles. Conversely, under high heating rates, the temperature distribution becomes significantly flatter, thereby validating the average temperature approximation. This assumption is further supported by finite element modeling and experimental observations[27], which show that for typical heater geometries—whether meander-like or strip-like—and sensing areas on the millimetre scale, temperature inhomogeneities remain within approximately 20 K at heating rates on the order of $10^5$ K/s.

In the quasi-adiabatic method, nanocalorimeters are driven by short pulses of constant current, sufficiently intense to induce heating rates that far exceed the characteristic cooling rate of the system. By locally measuring the voltage drop across the calorimetric cell using a four-wire configuration, one can determine the instantaneous power input as $P_{in}(t) = V(t) \cdot I(t)$, as well as the time-dependent resistance $R(t) = V(t)/I(t)$. Using a prior calibration of the resistance-temperature relationship, the temperature evolution $T(R(t))$ , can be retrieved, and consequently, the effective heating rate can be calculated as, $\beta(t) = dT(t)/dt$.

A straightforward analysis leads to the conclusion that the apparent heat capacity of a single calorimetric cell, $Cp_{\beta}^{cc}(t)$, can be expressed as:

$$Cp_{\beta}^{cc}(t) = \frac{P_{in}(t)}{\beta(t)} = Cp_{real}^{cc}(t) + \frac{P_{losses}(t)}{\beta(t)}$$

where $P_{losses}$ term includes contributions from thermal conduction through the membrane and radiative heat losses, and thus, accounts for the energy lost per unit time due to non-adiabatic effects. Power losses depend primarily on the temperature difference between the calorimetric cell and its surroundings, and also on the internal temperature distribution within the cell at a given time. To ensure quasi-adiabatic conditions, the input power $P_{in}$ must be significantly greater than $P_{losses}$, thereby minimizing the influence of thermal losses. As noted by Efremov et al. [1], a first-order estimation of $P_{losses}$ can be performed for each calorimetric cell by measuring the apparent heat capacity at different heating rates. The plot of $Cp_{\beta}^{cc}(T)$ as a function $1/\beta$ at a fixed temperature allows for the extrapolation of the real heat capacity $Cp_{real}^{cc}$ as the intercept, and the loss term $P_{losses}$ as the slope. Nevertheless, the approximation of adiabatic behaviour becomes valid for most practical scenarios - such as at high heating rates or within a narrow temperature scan range - the apparent heat capacity $Cp_{\beta}^{cc}$ closely matches the real heat capacity $Cp_{real}^{cc}$. In this framework, the measurement of a sample requires two consecutive measurements of the same calorimetric cell: first with the empty CC (denoted with superscript 0) and subsequently with the sample evaporated onto it (denoted with superscript 1).

$$C_P{}^{sample}\,(T) \approx C_{P_\beta}^{cc,1}(T) - C_{P_\beta}^{cc,0}(T)$$

Analysing the intrinsic noise of this measurement reveals that it is largely dominated by the inefficiency of computing the time derivative of a digitized signal. The apparent heat capacity is obtained by measuring the voltage drop across the calorimetric cell, $V_S$, and across a load resistor, $V_I$, which is used to determine the current supplied to the device $I(t)$. These signals are acquired using a 16-bit data acquisition system (DAQ) operating at a given sampling frequency $f$. In the best-case scenario when utilizing the full dynamic range of the DAQ, the base digital noise used to be equivalent to approximately three times the least significant bit (LSB), and then the signal-to-noise ratio (SNR) (defined as $SNR = V/\eta_V$, where $\eta_V$ is the root mean square (RMS) noise of the signal $V$) remains below 2x10$^4$.

The uncertainty in the input power term, $\eta_P$, in the expression for apparent heat capacity can be estimated as:

$$\eta_P = \sqrt{2} \cdot V \cdot \eta_V / R_{\text{load}},$$

assuming that $R_{load}$ is typically chosen such that $V_I \sim V_S$, and therefore both signals share a similar noise level $\eta_V$.

Similarly, considering the acquisition frequency $f$ and the temperature coefficient of resistance $\alpha \cdot R_o$, the propagated uncertainty in the heating rate measurement can be expressed as:

$$\eta_B = \sqrt{2} \cdot \frac{\eta_v \cdot R_{load}}{V \cdot R_o \cdot \alpha} \cdot \mathrm{f} \approx \frac{\sqrt{2} \cdot f}{SNR \cdot \alpha}$$

By rewriting the expression, it becomes evident that a higher signal-to-noise ratio (SNR) in the voltage measurement leads to a lower uncertainty $\eta_B$. However, the presence of the acquisition frequency $f$ in the numerator of the expression highlights the significant impact that differentiating digitized signals has on the uncertainty, thereby fundamentally limiting the precision of the method. A traditional strategy used in instrumentation to overcome this lack of precision is to consider differential measurements. Using a couple of twin calorimetric cells working simultaneously, differential signatures can be widely amplified while fitting the range of the ADC card. By this approach we can effectively reach the limit of the intrinsic thermal noise (or Johnson-Nyquist noise) of the metallic elements used as thermometer and with the pre-amplification stage.

The differential methodology can be implemented by connecting the nanocalorimeters in series and performing the differentiation using instrumentation amplifiers (as shown in Figure 2 (a) of the article).

The differential voltage drop between the two calorimetric cells, defined as $\Delta V = V_S - V_R$, where cell S hosts the sample and cell R serves as the reference, can be acquired with high amplification (typically ×10 to ×1000) using an instrumentation amplifier. The difference in apparent heat capacity between the two cells can then be recalculated by incorporating the differential voltage ΔV into the general expression:

$$\Delta C_P(T_S(t)) = \frac{P_S}{\beta_S} - \frac{P_R}{\beta_R}$$

Note that the differential heat capacity between both devices, $\Delta C_P(T_S(t))$, is a sample temperature function, which in turn is a time function. For convenience, this dependence is not shown explicitly.

As a first step we can express the power terms as function of the voltage dropped in the calorimetric cells:

$$\Delta C_P = \frac{V_S \cdot I}{\beta_S} - \frac{V_R \cdot I}{\beta_R} = \frac{I \cdot V_R}{\beta_R}\left(\frac{V_S \cdot \beta_R}{V_R \cdot \beta_S} - 1\right)$$

We use the differential configuration relation, $V_S = V_R + \Delta V$, to obtain

$$\Delta C_P = \frac{I \cdot V_R}{\beta_R}\left[\frac{(\Delta V + V_R)\cdot \beta_R}{V_R \cdot \beta_S} - 1\right] = \frac{I \cdot \Delta V}{\beta_S} + \frac{I \cdot V_R}{\beta_S}\left(1 - \frac{\beta_S}{\beta_R}\right)$$

As in the development of Efremov et al.[1] we spread the expression for $.d\Delta V/dt$:

$$\Delta V = V_S - V_R \rightarrow \frac{d\Delta V}{dt} = \frac{dV_S}{dt} - \frac{dV_R}{dt}$$

If functions $V(t)$ and $T(t)$ are derivative and continuous, then $dV/dt = (dV/dT)(dT/dt)$. Using this relation on previous equation) and by reordering we can write:

$$\frac{d\Delta V}{dt} = \frac{dV_S}{dT_S}\beta_S - \frac{dV_R}{dT_R}\beta_R \quad \rightarrow \quad \frac{\beta_S}{\beta_R} = \frac{(d\Delta V/dt)_t}{\beta_R\,(dV_S/dT_S)_t} + \frac{(dV_R/dT_R)_t}{(dV_S/dT_S)_t}$$

By substituting the equivalent expression for the ratio between heating rate, we obtain:

$$\Delta C_P(T_S(t)) = \frac{I \cdot \Delta V}{\beta_S} + \frac{I \cdot V_R}{\beta_S}\left[1 - \left(\frac{(d\Delta V/dt)_t}{\beta_R\,(dV_S/dT_S)_t} + \frac{(dV_R/dT_R)_t}{(dV_S/dT_S)_t}\right)\right]$$

And finally considering that current injected is constant in time, then $V_i = I_i\, R_i \longrightarrow d(V_i) = d(I_i\, R_i) = I_i\, d(R_i)$ and can be written in the expression that is routinely used in quasi-adiabatic experiments and has been adopted in microsecond pulsed experiments:

$$\Delta C_P(T_S(t)) = \frac{I\,\Delta V}{\beta_S} - \frac{V_R}{\beta_S\,\beta_R}\frac{(d\Delta V/dt)_t}{(dR_S/dT_S)_t} + \frac{V_R\, I}{\beta_S}\left[1 - \frac{(dR_R/dT_R)_t}{(dR_S/dT_S)_t}\right]$$